
\font\lbf=cmbx10 scaled\magstep2

\def\bs{\bigskip}
\def\ms{\medskip}
\def\np{\vfill\eject}
\def\nl{\hfill\break}
\def\ni{\noindent}
\def\cl{\centerline}

\def\title#1{\cl{\lbf #1}\ms}
\def\ctitle#1{\bs\cl{\bf #1}\par\nobreak\ms}
\def\stitle#1{\bs{\ni\bf #1}\par\nobreak\ms}

\def\ref#1#2#3#4{#1\ {\it#2\ }{\bf#3\ }#4\par}
\def\ns{\kern-.33333em}
\def\CQG{Class.\ Qu.\ Grav.}
\def\CMP{Comm.\ Math.\ Phys.}
\def\JMP{J.\ Math.\ Phys.}
\def\NP{Nucl.\ Phys.}
\def\PL{Phys.\ Lett.}
\def\PR{Phys.\ Rev.}
\def\PRL{Phys.\ Rev.\ Lett.}
\def\PRS{Proc.\ R.\ Soc.\ Lond.}

\def\abs#1{\left\vert#1\right\vert}
\def\k{\kappa}
\def\t{\theta}
\def\L{\Lambda}
\def\i{\imath}

\def\third{{\textstyle{1\over3}}}
\def\twothird{{\textstyle{2\over3}}}
\def\quart{{\textstyle{1\over4}}}

\def\eg{{\it e.g.\ }}
\def\ie{{\it i.e.\ }}
\def\cf{{\it cf\/\ }}

\def\etal{{\it et al\/\ }}

\magnification=\magstep1

\title{Global structure of a black-hole cosmos and its extremes}
\ctitle{Dieter R. Brill* and Sean A. Hayward}
\cl{Max-Planck-Institut f\"ur Astrophysik}
\cl{Karl-Schwarzschild-Stra\ss e 1}
\cl{8046 Garching bei M\"unchen}
\cl{Germany}
\ms\cl{* Permanent address: Department of Physics}
\cl{University of Maryland}
\cl{College Park MD 20742}
\cl{U.S.A.}
\bs\cl{6th April 1993}
\bs\cl{Submitted to {\it\CQG}}
\bs\ni{\bf Abstract.}
We analyze the global structure of a family of Einstein-Maxwell solutions
parametrized by mass, charge and cosmological constant.
In a qualitative classification there are:
(i) generic black-hole solutions, describing a Wheeler wormhole
in a closed cosmos of spatial topology $S^2\times S^1$;
(ii) generic naked-singularity solutions,
describing a pair of ``point" charges in a closed cosmos;
(iii) extreme black-hole solutions, describing a pair of ``horned"
particles in an otherwise closed cosmos;
(iv) extreme naked-singularity solutions,
in which a pair of point charges forms and then evaporates,
in a way which is not even weakly censored;
and (v) an ultra-extreme solution.
We discuss the properties of the solutions and of various coordinate systems,
and compare with the Kastor-Traschen multi-black-hole solutions.
\ms
\stitle{I. Introduction}\ni
In classical gravity there is a family of black-hole solutions
characterized by mass $M$, charge $Q$ and cosmological constant $\L$
(Carter 1973).
Such solutions are often considered most realistic,
for example for astrophysical applications,
when the black hole is of a generic type. However,
today renewed interest is attached to various types of {\it extreme} objects
(see for example Banks \etal 1992a).
Additionally, typical inflationary cosmologies have an appreciable $\L$
during some epoch.
For such reasons, it is of interest to examine cosmological black-hole
spacetimes\footnote\dag
{A spacetime with future null infinity $\Im^+$ contains black holes if
the causal past of $\Im^+$ is not the whole spacetime. We call a black-hole
spacetime cosmological if it has spatial sections that are compact
and bounded only by the black holes' horizon,
namely the boundary of the causal past of $\Im^+$.
The spacetimes we consider are
time-reversal-symmetric, so that
each ``hole'' has both black and white horizons,
but we follow convention in discriminating against the white holes.}
for their global properties and to identify the various extreme objects that
can be obtained by suitable choice of the parameters $(M,Q,\L)$.

Furthermore, Kastor and Traschen (1992) have recently given the first
cosmological Einstein-Maxwell solution containing many black holes,
but in coordinates that do not cover the spacetime completely.
The solutions are parametrized by the cosmological constant $\L$,
$n$ masses $M_i$ and positions $(x_i,y_i,z_i)$,
with charges $\abs{Q_i}=M_i$.
In the case $\L=0$, they reduce to
the Majumdar-Papapetrou multi-black-hole solutions
(Hartle and Hawking 1972).
Understanding the global structure of the single-mass case,
and the ``cosmological'' coordinates used,
is a first step to understanding the global structure
of the general Kastor-Traschen solutions.

In Section II we recall briefly the conformal treatment of infinity
and a procedure for analytic continuation
of a large class of time-independent metrics.
In Section III we use this procedure to classify the $(M,Q,\L)$ solutions
according to their qualitative global structure.
In Section IV we discuss the properties of the spacetimes,
and in Section V we consider alternative coordinates.
The Conclusion includes an outline of the corresponding results
for the Kastor-Traschen solutions.

\stitle{II. Conformal infinity and analytic continuation}\ni
The $(M,Q,\L)$ spacetimes have infinite regions whose physical significance
requires careful analysis.
One such region is the late time unbounded inflationary era, if $\L>0$. Also,
in some of these spacetimes the black holes are spatially infinite ``horns''.
Null geodesics may or may not cross such infinite regions in finite
affine parameter intervals (Carter 1973).
A convenient way to analyze such infinities and their global structure is
to change the metric by an unbounded conformal factor (Penrose 1963, 1965).
Surfaces at infinity become conformally finite,
and the global structure of certain simple spacetimes can be summarized
in conformal diagrams (\eg Hawking and Ellis 1973).
Apart from conformal infinity, the other key causal features are
the various horizons that occur in such spacetimes,
such as black-hole event horizons and cosmological horizons.

Originally, analyses of global structure were performed separately
for such examples as the Schwarzschild solution
(Finkelstein 1958, Kruskal 1960),
the Reissner-Nordstr\"om solution (Graves and Brill 1960, Carter 1966b)
and the Kerr solution (Carter 1966a, Boyer and Lindquist 1967).
Walker (1970) has given a general prescription which can be applied to
any spacetime of the type ${\cal T}\times{\cal S}$,
where ${\cal T}$ is
totally geodesic with time-independent metric of the form
$$ds^2=-F\,dT^2+F^{-1}dR^2,\qquad F=F(R),\eqno(1)$$
where the geometry of ${\cal S}$
is non-singular and a smooth function of $R$,
and where $F$ has one or more simple zeros,
which correspond to Killing horizons (Carter 1969).
Namely, if $R_1$ and $R_2$ are two successive zeros of $F$,
then the conformal diagram of the region
$(T,R)\in(-\infty,\infty)\times(R_1,R_2)$
is a ``block'' with four inner null boundaries corresponding to
$(T,R)=(\pm\infty,R_i)$.
Similarly, if $R_1$ is the highest zero of $F$, then the region
$(T,R)\in(-\infty,\infty)\times(R_1,\infty)$
is a block with an outer conformal boundary.
Blocks are glued together at the inner null boundaries
so that $R$ becomes a smooth (or analytic) function on the total space,
and $T$ has a logarithmic singularity at the boundary.
Once two or more blocks are glued together, further extensions can
also be generated by noting that the symmetry $T\mapsto-T$ of
each block must extend to a symmetry of the analytic extension.

It can be shown that coordinate charts exist which cover the boundaries,
as follows.
Denote the simple zero by $R=R_0\not=0$,
so that $F(R_0)=0$, $F'(R_0)=2\k\not=0$,
with $\abs\k$ being the surface gravity of the Killing horizon.
Then Kruskal-like coordinates
$$u=-e^{\k(R^*-T)},\qquad
v=e^{\k(R^*+T)},\qquad
R^*=\int F^{-1}dR\eqno(2)$$
can be introduced, in terms of which the line-element (1) becomes
$$ds^2=-\k^{-2}Fe^{-2\k R^*}du\,dv+R^2dS^2,\eqno(3)$$
where $R^*=(2\k)^{-1}\log(-uv)$,
and $R$ and $F$ are implicitly determined as functions of $uv$.
Near $R=R_0$,
$$F=2\k(R-R_0)+O(R-R_0)^2,\qquad
R^*=(2\k)^{-1}\log(R-R_0)+O(R-R_0),\eqno(4)$$
where the constant of integration has been fixed, and it follows that
$$ds^2=-2\k^{-1}du\,dv+R_0^2dS^2+O(R-R_0),\eqno(5)$$
which is manifestly non-singular.
Explicit coordinate charts are given in Section V for the case of interest.

The case of coincident roots can be treated by a simple extension of
this procedure: when two blocks are glued across a double root,
which corresponds to a degenerate Killing horizon
in the sense of Carter (1973),
$T$ has a simple pole at the null boundary,
and the usual spatio-temporal inversion of $(T,R)$ does not occur.
A double (or higher multiple) root of $F$ also implies that
the geodesic distance along a constant-$T$ surface from
any point within the block to the degenerate horizon is infinite.
If $T$ is a timelike coordinate this means that the horizon is at an infinite
distance, and the spatial geometry has the shape of a ``horn''
rather then a ``wormhole''.

\stitle{III. Global structure}\ni
The $(M,Q,\L)$ solutions (Carter 1973) have line-element
$$ds^2=-F\,dT^2+F^{-1}dR^2+R^2dS^2,\eqno(6)$$ where
$$F(R)={{Q^2}\over{R^2}}-{2M\over{R}}+1-\third\L R^2\eqno(7)$$
and $dS^2$ refers to the unit 2-sphere.
The case $\L=0$ yields the Reissner-Nordstr\"om solution,
which was analyzed by Graves and Brill (1960) and Carter (1966b).
The case $Q=0$ corresponds, for $9\L M^2<1$,
to a Schwarzschild black hole in a de Sitter cosmos,
which was analyzed by Gibbons and Hawking (1977).
Henceforth we assume that $\L>0$, $M>0$ and $Q\not=0$.
The cases $\L<0$ and $M<0$ can be analyzed similarly,
but are of less interest for either cosmology or black holes.

According to the procedure explained in the previous Section,
the global structure is determined by:
behaviour at $R=0$, where there is a scalar curvature singularity;
behaviour at $R=\infty$,
which for $\L>0$ is both timelike infinity $\i^{\pm}$
and null infinity $\Im^{\pm}$;
and behaviour at the Killing horizons $F(R)=0$.
It is easily checked that the quartic $R^2F(R)$ has exactly one negative root,
and generically either one or three positive roots,
with special cases of double or triple roots.
The possibilities for the positive roots are as follows.
\item{(i)} The generic black-hole case: three simple roots $R_c>R_o>R_i$.
There are three types of Killing horizon:
cosmological horizons at $R=R_c$ and inner and outer black-hole horizons
at $R=R_i$ and $R=R_o$ respectively.
\item{(ii)} The generic naked-singularity case: one simple root,
giving only cosmological horizons.
\item{(iii)} The extreme (or ``cold'') black-hole case:
a double root $R_d$ and a simple root $R_c>R_d$,
corresponding to a degenerate black-hole horizon at $R=R_d$
and a cosmological horizon at $R=R_c$.
\item{(iv)} The extreme (or marginal) naked-singularity case:
a double root $R_d$ and a simple root $R_i<R_d$,
corresponding to a degenerate cosmological horizon at $R=R_d$
and an inner horizon at $R=R_i$.
\item{(v)} The ``ultra-extreme'' case:
triple root, with one doubly degenerate horizon.
\par\ni
Figure 1 shows the conformal diagrams as obtained by the method of Section II.
Many properties of the solutions are clear from these diagrams,
and we make additional comments in the next Section.

To make explicit which case occurs for each $(M,Q,\L)$, we note
that the double roots occur if and only if
$M=M_\pm(Q,\L)$, where
$$M_\pm=P_\pm(1-\twothird\L P_\pm^2),\qquad
P_\pm^2={1\over{2\L}}\left(1\pm\sqrt{1-4\L Q^2}\right),\eqno(8)$$
with the double root being at $R=P_\pm$ (Romans 1992).
The ultra-extreme case occurs when $9M^2=8Q^2=2/\L$, so that $P_+=P_-$.
Otherwise, the extreme black hole occurs if $M=M_-(Q,\L)$,
and the extreme naked singularity occurs if $M=M_+(Q,\L)$.
The generic black hole occurs if $M_-(Q,\L)<M<M_+(Q,\L)$,
and the generic naked singularity otherwise.
Figure 2 shows where the various cases lie in parameter space.

As an example of relevance to the Kastor-Traschen solutions, we note that
for the case $\abs{Q}=M$,
a generic black hole occurs if $\L M^2<3/16$,
an extreme naked singularity if $\L M^2=3/16$,
and a generic naked singularity if $\L M^2>3/16$.
The extreme black hole does not occur in this class.
If $\L M^2 \ll 1$
the Killing horizons are located approximately at
$$R_i\sim M-HM^2,\qquad R_o\sim M+HM^2,\qquad R_c\sim H^{-1}-M,\eqno(9)$$
where $H=\sqrt{\L/3}$.
Romans (1992) gives such expressions to a higher order of approximation.

\stitle{IV. Properties}\ni
The generic black-hole solution, Figure 1(i), can be interpreted as
a pair of oppositely charged Reissner-Nordstr\"om black holes
at opposite poles of a de Sitter cosmos (Mellor and Moss 1989).
The interiors of the holes lead either to other de Sitter regions,
or to each other, depending on the topological identifications made.
We refer to the regions $R<R_o$ as the holes,
and the regions $R>R_c$ as the cosmological regions.
Any constant-$T$ surface is a surface of reflection symmetry. When
$F<0$ this surface is Lorentzian and corresponds to reflection about the
``equator" of the de Sitter exterior, or about the ``throat" of the holes.
For $F>0$, $R_o<R<R_c$, the surfaces are free from singularities and
can be connected smoothly to form
a surface of time symmetry.
This surface divides the spacetime into
a future containing a black hole ($R<R_o$)
and an expanding cosmological region ($R>R_c$),
and a past containing a white hole and a contracting cosmological region.

The spatial geometry on such constant-$T$ surfaces is given by
$$F^{-1}dR^2+R^2dS^2,\eqno(10)$$
which can be embedded in four-dimensional flat space
$$dZ^2+dR^2+R^2dS^2\eqno(11)$$ by
$$Z=\int\sqrt{F^{-1}-1}\,dR,\eqno(12)$$ as depicted in Figure 3.
The surface has maximal radius $R_c$ at the cosmological horizon,
and minimal radius $R_o$ at the throats of the holes.
Isometric copies of the surface can be smoothly joined at the throats,
producing a periodic $S^2\times R^1$ spatial topology,
in which the interiors of the black holes lead to further de Sitter regions.
Alternatively, one may simply identify the throats with each other,
giving an $S^2\times S^1$ spatial topology.
The latter identification provides a classic example
of a charged wormhole in the sense of Wheeler (1962):
the lines of electric force converge on one side of the wormhole
and diverge from the other, so that the appearance of two opposite charges
is generated by non-trivial topology rather than by charged sources.
This appears to be the only explicitly known example of a Wheeler wormhole
spacetime without ``external'' fields.

The generic naked-singularity solution is shown in Figure 1(ii).
A constant-$T$ spatial surface has topology $S^2\times R^1$,
which can be visualized as an $S^3$ punctured at opposite poles,
with the two singularities being at finite affine distance.
The lines of electric force diverge from one pole and converge on the other,
so that the singularities have the appearance of point charges.

We now turn to the extreme cases.
The extreme black-hole solution, Figure 1(iii), can be interpreted as
a pair of oppositely charged extreme Reissner-Nordstr\"om black holes
at opposite poles of a de Sitter cosmos.
Note that for $\L>0$ the extreme case occurs for $M<\abs{Q}$.
That is, in the cosmological context, the maximal charge
on a black hole---beyond which there is a naked singularity instead---is
larger than in asymptotically flat space (Romans 1992, \cf Banks \etal 1992b).
The topology of a constant-$T$ spatial surface is $S^2\times R^1$.
The two minimal-radius spheres, which also represent the black holes'
horizons, are located at infinite affine distance on this spatial surface,
preventing a wormhole identification, so that these black holes
are horns rather than wormholes.

The extreme naked-singularity case, Figure 1(iv), has a novel structure
that suggests the creation and subsequent annihilation
of a pair of point charges.
A constant-$T$ spatial surface has topology $S^2\times R^1$,
which again can be visualized as an $S^3$ punctured at opposite poles,
with the two singularities being at finite affine distance,
and the lines of electric force
diverging from one pole and converging on the other.
Unlike the generic naked singularities,
or that of the negative-mass Schwarzschild solution,
these singularities do not exist for all time,
but develop from an initially regular state,
\ie there are partial Cauchy surfaces,
of topology $S^2\times R^1$ or $S^2\times S^1$,
depending on the identifications made.
The singularities are not even weakly censored, in the sense that
any observer whose future life is long enough will see them:
any path from $\i^-$ to $\i^+$ passes through the causal future
(and the causal past) of the singularities.
This is in distinction to the generic black hole,
or the Reissner-Nordstr\"om black hole,
whose singularities remain unseen by wary observers.
Another novel feature is that the singularities subsequently dissolve,
with the process being visible from $\Im^+$.

Finally, we note that the ultra-extreme case
is similar to the generic naked-singularity case,
except that the singularities are at infinite affine distance.

\stitle{V. Other charts}\ni
We now consider alternative coordinate systems, for three main reasons.
Firstly, the static coordinates $(T,R)$ break down at the Killing horizons,
and while it was shown in Section II that coordinate charts do exist there,
it is useful to find such charts explicitly.
Secondly, the Kastor-Traschen solutions are given in different coordinates,
and understanding the nature of these ``cosmological coordinates''
is important for the physical interpretation.
Thirdly, different coordinates emphasize different features, which may
or may not persist in the general Kastor-Traschen solutions.

As a preliminary example, we consider the de Sitter cosmos
(Hawking and Ellis 1973).
It is completely covered by the usual ``hyperbolic'' de Sitter coordinates,
$$ds^2=3\L^{-1}(-d\t^2+\cosh^2\t\,d\Sigma^2),\eqno(13)$$
where $d\Sigma^2=d\chi^2+\sin^2\chi\,dS^2$ refers to the unit 3-sphere.
The conformal diagram, Figure 4(i),
shows conformal infinity $\Im^\pm$ ($\t=\pm\infty$)
and the (opposite but otherwise arbitrary) poles of the 3-sphere,
$\chi=0$ and $\chi=\pi$.
In such diagrams one can also show other coordinates and the extent to
which they cover the complete spacetime. Figure 4(ii) shows this for the
``cosmological'' (or ``steady-state-universe'')
coordinates $(t_+,r)$ or $(t_-,r)$, in terms of which the de Sitter metric is
$$ds^2=-dt_\pm^2+e^{\pm2Ht}(dr^2+r^2dS^2),\eqno(14)$$
where the coordinate transformation is
$$\pm Ht_\pm=\log(\sinh\t+\cosh\t\cos\chi),\qquad
\pm Hr={\cosh\t\sin\chi\over{\sinh\t+\cosh\t\cos\chi}},\eqno(15)$$
and $H=\sqrt{\L/3}$.
We see that most of de Sitter spacetime is covered by two charts,
one ``expanding'' $(t_+,r)$ chart and one ``contracting'' $(t_-,r)$ chart,
but that both fail to cover the cosmological horizon $t_\pm=\mp\infty$.
By the reflection $\t\mapsto-\t$
one obtains similar coordinates that cover the previous cosmological horizon
but break down at the reflected cosmological horizon.

The line-element may also be written in static $(T,R)$ coordinates,
$$ds^2=-F\,dT^2+F^{-1}dR^2+ R^2dS^2, \qquad F=1-\third\L R^2,\eqno(16)$$
which are related to the cosmological coordinates by
$$R=e^{\pm Ht_\pm}r,\qquad t_\pm=T\mp\int HRF^{-1}dR,\eqno(17)$$
and are shown in Figure 4(iii).
The static coordinates divide de Sitter spacetime into four regions,
and break down at both cosmological horizons. The two regions that
do not include $\Im^{\pm}$ are static in the usual sense.
The regions attached to $\Im^+$ and $\Im^-$
are expanding and contracting respectively, with constant-$T$ surfaces
being spatial homogeneous cylinders.
Thus the cosmological coordinates have an advantage
over the static coordinates,
namely that they can be used to cover more of the manifold,
but have the disadvantage that by breaking the time symmetry
they can give a somewhat misleading picture of the spacetime.\footnote\ddag
{The Schwarzschild solution is a similar example.
The static Schwarzschild coordinates break down at the horizons,
which was originally taken as indicating
a singularity or boundary of the spacetime.
Finkelstein (1958) showed that
the black-hole horizon could be covered by different coordinates,
but concluded that this indicated a time asymmetry of the spacetime.
Kruskal (1960) showed how to cover the whole manifold with a different chart.}

The $(M,Q,\L)$ solutions (6)--(7)
can likewise be written in cosmological (or isotropic) coordinates
$(t_+,r)$ or $(t_-,r)$,
$$ds^2=-V^{-2}dt_\pm^2+U^2e^{\pm2Ht_\pm}(dr^2+r^2dS^2),\eqno(18)$$
where
$$U(\rho)=1+M\rho^{-1}+\quart(M^2-Q^2)\rho^{-2},\qquad
V(\rho)={U(\rho)\over{1-\quart(M^2-Q^2)\rho^{-2}}}\eqno(19)$$
and
$$\rho=e^{\pm Ht_\pm}r,\eqno(20)$$
with $H=\sqrt{\L/3}$ as before. Here the relation between the coordinates is
$$R=U\rho,\qquad t_\pm=T\mp\int{HR\,dR\over{F\sqrt{F+H^2R^2}}}.\eqno(21)$$
These coordinates yield various charts (Figure 5)
which cover several of the static blocks for $R>R_+$ or $R<R_-$,
but with a gap for $R_-\le R\le R_+$, where
$$R_\pm=M\pm\sqrt{M^2-Q^2}.\eqno(22)$$
Because $\rho$ is not a single-valued function of $R$,
successive constant-$t$
surfaces intersect along a caustic at $R_\pm$,
and the constant-$r$ surfaces have cusps there,
with the two branches corresponding to different charts
related by $T\mapsto-T$.
The black-hole horizon occurs at $\pm HVU\rho=1$.

For the case $\abs{Q}=M$, the gap between $R_\pm$ disappears and the solution
(18)--(20) coincides with
the single-mass Kastor-Traschen solution in cosmological coordinates.
A further transformation to
$$\tau_\pm=\pm H^{-1}e^{\pm Ht_\pm}\eqno(23)$$
puts the line-element in the form
$$ds^2=-(Mr^{-1}\pm H\tau_\pm)^{-2}d\tau_\pm^2
+(Mr^{-1}\pm H\tau_\pm)^2(dr^2+r^2dS^2).\eqno(24)$$
Now a single $(\tau,r)$ chart covers,
for $\tau<0$ and $\tau>0$, two $(t,r)$ charts,
whereas the latter break down at $\tau=0$, corresponding to $R=M$.
Figure 6 shows the new cosmological coordinates for the $\abs{Q}=M$ solution.
A single $(\tau,r)$ chart covers four $(T,R)$ charts,
including the three Killing horizons between them,
namely the cosmological, event and inner horizons.
For $(\tau_+,r)$,
this includes the expanding cosmological region and part of the white hole,
while for $(\tau_-,r)$,
the contracting cosmological region and part of the black hole are covered.
The boundaries of the $(\tau_+,r)$ chart are the singularity $Hr\tau_+=-M$,
the other event horizon $(\tau_+,r)=(\infty,0)$,
the other cosmological horizon $(\tau_+,r)=(0,\infty)$,
the other inner horizon $(\tau_+,r)=(-\infty,0)$,
and $\Im^+$ ($\tau_+=\infty$, $r$ finite).
By using the time-inversion symmetry $T\mapsto-T$ of the static frame,
similar cosmological charts can be constructed to cover the entire manifold,
except for the isolated 2-surfaces where the Killing horizons cross.
These are regular since the constant-$T$ surfaces are smooth
and totally geodesic (Section IV).

Thus we have verified the global structure derived in Section III.
The cosmological coordinates are useful for covering the horizons,
but give highly distorted charts which do not respect the time symmetry.
The time symmetry can be restored by yet another transformation,
to double-null coordinates $(u,v)$.
First note that the $\abs{Q}=M$ case
can be written in $(t_\pm,\rho)$ coordinates as
$$ds^2=-U^{-2}dt_\pm^2+U^2(d\rho\mp H\rho dt_\pm)^2+U^2\rho^2dS^2,\eqno(25)$$
where $U(\rho)=1+M\rho^{-1}$. Then the transformation
$$u=t_\pm+\int{d\rho\over{U^{-2}\mp H\rho}},\qquad
v=t_\pm-\int{d\rho\over{U^{-2}\pm H\rho}}\eqno(26)$$
puts the line-element in the form
$$ds^2=-(U^{-2}-\third\L U^2\rho^2)du\,dv+U^2\rho^2dS^2,\eqno(27)$$
where $\rho$ is implicitly determined as a function of $u-v$ by (26).
This double-null form has the manifest time-inversion symmetry
$$(u,v)\mapsto(-v,-u),\eqno(28)$$
which allows the recovery of all the different $(t,r)$ or $(\tau,r)$ charts.

Although the cosmological coordinates cover some horizons of the
spacetimes discussed here, others occur at $\tau = 0$ and $\tau = \infty$
in these coordinates. Some cosmological models take a metric like
(18) or (24) to be valid only for a finite range of $\tau$.
The global structure  of such models can be quite different from
that shown in our Figures.  For example, if the universe started
out as an expanding $k=0$ Friedmann model with $\Lambda = 0$, and
$\Lambda$ was ``turned on" at some finite $\tau$, then $\Im^-$ in Figure 6
would be replaced by the big bang singularity.

\stitle{VI. Conclusion}\ni
We have seen that the metrics that appear to describe
a single charged black hole with cosmological constant
may be maximally extended to a spacetime
that generically has an (approximately de Sitter) cosmological region
and {\it two} black hole regions for each $\Im^+$.
In the alternative extension with $S^2\times S^1$ spatial topology,
there is only one $\Im^+$ and one black hole,
with the black hole having two horizons at opposite poles of the cosmos,
constituting a Wheeler wormhole.
In addition, there is a novel type of extreme case
where a pair of point charges forms and subsequently dissolves.

The cosmological or isotropic coordinates are simplest in the case
$\abs{Q}=M$, where isometric copies of a single coordinate chart suffice
to cover most of the maximal extension.
The ``Hubble constant'' $\pm H$ is positive in some of these charts
and negative in others,
though the adjectives ``expanding'' and ``contracting''
can really be justified
only for the blocks attached to $\Im^+$ and $\Im^-$ respectively.
Each cosmological chart breaks down at a cosmological horizon
given by a finite value of the coordinate $\rho$ of (20),
namely the largest solution of $\pm HVU\rho=1$,
approximately $\rho\sim\pm H^{-1}-2M$.
This horizon is covered by a different cosmological chart
obtained using the time-inversion symmetry (28).

The Kastor-Traschen metric is similar to (24), with $Mr^{-1}$
replaced by $\sum_i m_i r_i^{-1}$, where $r_i$ are Euclidean distances
of the field point $r$ from $n$ fixed centers, and $m_i$ are mass
parameters associated with these centers.
At large $\tau$ the metric approaches that of (24),
so that $\Im^{\pm}$ exist as for the single-mass case.
The incompleteness at large $r$ is also similar to that
of the single-mass case, and the extension
across the horizon at $(\tau, r) = (0, \infty)$ leads to a symmetrically
related spacetime
that contains the opposite charge. In the Kastor-Traschen metric
this region is described by an identical expression but with the
opposite sign of $H$. Thus in a background de Sitter space,
we have a set of black holes with one sign of charge joined across
the horizon to a corresponding set with the opposite charge.
Details of this geometry as well as the continuation across the
black-hole horizons will be discussed in a separate paper.

\bs\ni
Acknowledgements.
We would like to thank Piotr Chru\'sciel, David Kastor and Jennie Traschen
for discussions,
and the Max-Planck-Gesellschaft for financial support.
\np
\begingroup
\parindent=0pt\everypar={\global\hangindent=20pt\hangafter=1}\par
{\bf References}\ms
\ref{Banks T, Dabholker A, Douglas M R, \& O'Loughlin M 1992a}\PR{D45}{3607}
\ref{Banks T, O'Loughlin M \& Strominger A 1992b}
{Black-hole remnants and the information puzzle}\ns{(hep-th/9211030)}
\ref{Boyer R H \& Lindquist R W 1967}\JMP8{265}
\ref{Carter B 1966a}\PR{141}{1242}
\ref{Carter B 1966b}\PL{21}{423}
\ref{Carter B 1969}\JMP{10}{70}
\ref{Carter B 1973 in}{Black Holes}\ns
{ed DeWitt C \& DeWitt B S (Gordon \& Breach)}
\ref{Finkelstein D 1958}\PR{110}{965}
\ref{Gibbons G W \& Hawking S W 1977}\PR{D15}{2738}
\ref{Graves J C \& Brill D R 1960}\PR{120}{1507}
\ref{Hartle J B \& Hawking S W 1972}\CMP{26}{87}
\ref{Hawking S W \& Ellis G F R 1973}{The Large-Scale Structure of Space-Time}
\ns{(Cambridge University Press)}
\ref{Kastor D \& Traschen J 1992}
{Cosmological multi-black-hole solutions}\ns{(hep-th/9212035)}
\ref{Kruskal M D 1960}\PR{119}{1743}
\ref{Mellor F \& Moss I 1989}\PL{B222}{361}
\ref{Penrose R 1963}\PRL{10}{66}
\ref{Penrose R 1965}\PRS{A284}{159}
\ref{Romans L 1992}\NP{B383}{395}
\ref{Walker M 1970}\JMP{11}{2280}
\ref{Wheeler J A 1962}{Geometrodynamics}\ns{(Academic Press)}
\endgroup

\np
\stitle{Figure captions}\ni
1. Penrose-Carter conformal diagrams of the $(M,Q,\L)$ solutions, showing
\item{(i)} the generic black-hole case,
\item{(ii)} the generic naked-singularity case,
\item{(iii)} the extreme black-hole case,
\item{(iv)} the extreme naked-singularity case.
\par\ni
The conformal diagram for the ultra-extreme case
is identical to (ii).
Wavy lines represent $R=0$ singularities,
diagonal lines represent $F(R)=0$ Killing horizons,
and horizontal lines represent conformal infinity
$\Im^+$ or $\Im^-$ ($R=\infty$).
The curves represent surfaces of constant $R$.
The maximal analytic extension is obtained by identifying
isometric copies along the notched horizons.
\nl
{\it To be inserted in Section III.}
\ms\ni
2. Parameter space $(M,Q)$ for fixed $\L>0$.
Extreme black holes lie along the curve $M=M_-(Q)$,
and extreme naked singularities along $M=M_+(Q)$,
with the two curves meeting in a cusp,
corresponding to the ultra-extreme case.
Generic black holes lie inside the double curve,
and generic naked singularities outside.
\nl
{\it To be inserted in Section III.}
\ms\ni
3. Geometry of a constant-$T$ surface of time symmetry
for the generic black hole,
embedded in flat space, with a polar angle suppressed.
Here $HM=0.05$, $(HQ)^2=0.001$.
The curves represent lines of electric force.
The two wormhole throats can be smoothly identified to form
a Wheeler wormhole universe at a moment of time symmetry.
\nl
{\it To be inserted in Section IV.}
\ms\ni
4. Conformal diagrams of the de Sitter cosmos,
showing conformal infinity $\Im^{\pm}$
and the regular poles $O$.
The curves represent 3-surfaces of constant
(i) hyperbolic $(\t,\chi)$, (ii) cosmological $(t,r)$ and (iii) static $(T,R)$.
\nl
{\it To be inserted in Section V.}
\ms\ni
5. Conformal diagram as in Figure 1(i),
showing the cosmological, or isotropic, coordinates for the case $\abs{Q}<M$.
The curves represent 3-surfaces of constant $\tau_+$ and of constant $r$.
(Here $\tau_+$ is the coordinate defined in (23) and can take negative
values.)
Charts are shown for $R>R_+$ and $R<R_-$,
with the region $R_-\le R\le R_+$ not being covered by such charts.
The dashed curves show the continuation of one curve of constant
$\tau$, and one of constant $r$, beyond the caustic at $R=R_+$.
\nl
{\it To be inserted in Section V.}
\ms\ni
6. A ``cosmological'' $(\tau_+,r)$ chart
for the $\abs{Q}=M$ solution, for (i) a generic black hole,
(ii) the extreme naked singularity, and (iii) a generic naked singularity.
The curves represent surfaces of constant $\tau_+$ and $r$.
\nl
{\it To be inserted in Section V.}

\bye